\documentclass{llncs}

\usepackage{xspace}%
\usepackage{graphicx}%
\usepackage{subfigure}%
\usepackage{listings}%
{\description \item[\fbox{To Do}]~\sl%
  \typeout{----------- Something left to be done here!
    ------------------}%
}%
{\enddescription}%

\newcommand{\myblock}[1]{\mbox{\lstinline{#1}}}
\newcommand{\myblockxspace}[1]{\myblock{#1}\xspace}

\newcommand{\Offset}{\myblockxspace{offset}}
\newcommand{\Size}{\myblockxspace{size}}
\newcommand{\Datasize}{\myblockxspace{data_size}}
\newcommand{\Adjustedsize}{\myblockxspace{adjusted_size}}
\newcommand{\Blockid}{\myblockxspace{block_id}}
\newcommand{\Blockversion}{\myblockxspace{block_version}}
\newcommand{\Leftkey}{\myblockxspace{left_key}}
\newcommand{\Rightkey}{\myblockxspace{right_key}}
\newcommand{\Pageid}{\myblockxspace{page_id}}
\newcommand{\Providerid}{\myblockxspace{provider_id}}

\newcommand{\MB}{\mbox{MB}}

\begin{document}

\title{Distributed Management of Massive Data:\\
  an Efficient Fine-Grain Data Access Scheme%
%%  \thanks{Contact person: Gabriel Antoniu, IRISA, Campus de Beaulieu,
%%   F-35042 Rennes Cedex, France,
%%  \protect\url{Gabriel.Antoniu@inria.fr}.}%
}

%\titlerunning{an Efficient Fine-Grain Data Access Scheme%}

\author{%
  Bogdan Nicolae\inst{1}%
  \and%
  Gabriel Antoniu\inst{2}%
  \and%
  Luc Boug\'e\inst{3}%
}

%\authorrunning{Bogdan Nicolae et al.}

\tocauthor{Bogdan Nicolae, Gabriel Antoniu, Luc Boug\'e}

\institute{
  University of Rennes~1/IRISA, Campus de Beaulieu, 35042 Rennes Cedex, France\\\email{bogdan.nicolae@inria.fr}\\
  \and
  INRIA/IRISA, Campus de Beaulieu, 35042 Rennes Cedex, France\\\email{gabriel.antoniu@inria.fr}\\
  \and
  ENS Cachan Brittany/IRISA, Campus de Beaulieu, 35042 Rennes Cedex, France\\\email{luc.bouge@bretagne.ens-cachan.fr}
}

\maketitle

\begin{abstract}
  This paper addresses the problem of efficiently storing and
  accessing massive data blocks in a large-scale distributed
  environment, while providing efficient fine-grain access to data
  subsets. This issue is crucial in the context of applications in the
  field of databases, data mining and multimedia.  We propose a data
  sharing service based on distributed, RAM-based storage of data,
  while leveraging a DHT-based, natively parallel metadata
  management scheme. As opposed to the most commonly used grid storage
  infrastructures that provide mechanisms for \emph{explicit} data
  localization and transfer, we provide a \emph{transparent} access
  model, where data are accessed through global identifiers. Our
  proposal has been validated through a prototype implementation whose
  preliminary evaluation on the Grid'5000 testbed provides promising
  results.
\end{abstract}

%%%%%%%%%%%%%%%%%%%%%%%%%%%%%%%%%%%%%%%%%%%%%%%%%%%%%%%%%%%%%%

\section{Introduction}
\label{sec:intro}

Managing data at a large scale is paramount nowadays. Governmental and
commercial statistics, climate modeling, cosmology, genetics, 
bio-informatics, etc.  are just a few examples 
of fields routinely generating huge amounts of data. It becomes crucial to 
efficiently manipulate these data, which must be shared at the global scale.
In such a context, one important goal is to provide mechanisms
allowing to manage massive data blocks (e.g., of several terabytes),
while providing efficient fine-grain access to small parts of the
data. Several types of applications exhibit such a need for efficient
scaling to huge data sizes:
databases~(\cite{DouDou03PostgreSQL,Tho98Concurrency,NicJar00IEEE}), data
mining~\cite{JinYan05DataMining}, multimedia~\cite{casKur07Multimedia}, etc.

% An example of application is a very large database using a single file
% to store its whole data and metadata PostgreSQL~\cite{DouDou03PostgreSQL} and
% MySQL~\cite{mysql} use such an implementation scheme. Data mining
% applications provide another exampe of such a usage:...(to be
% developed).

\paragraph {Towards transparent management of data on the grid. }

The management of massive data blocks naturally requires the use of
data fragmentation and of distributed storage. 
Grid infrastructures, typically built by aggregating distributed
resources that may belong to different administration domains, 
provide an appropriate solution. When considering the existing
approaches to grid data management, we can notice that most of them
heavily rely on \emph{explicit} data localization and on
\emph{explicit} transfers of large amounts of data across the
distributed architecture: GridFTP~\cite{AllBesBreetal02},
Reptor~\cite{KunLauStoSto05FGCS}, Optor~\cite{KunLauStoSto05FGCS},
LDR~\cite{ldr},Chirp~\cite{Chirp}, IBP~\cite{BasBecFagetal02IBP}, NeST~\cite{BenVenLeRetal02},
etc. Managing huge amounts of data in such an explicit way at a very
large scale makes  the design of grid application much more complex.  One
key issue to be addressed is therefore the \emph{transparency} with
respect to data localization and data movements. Such a transparency
is highly suitable, as it liberates the user from the need to handle
data localization and transfers.

% As usually large amounts of input and output data need to be moved
% among clients and computing servers, most approaches to grid data
% management rely on file transfers. GridFTP~\cite{AllBesBreetal02},
% Chirp~\cite{Chirp} are typical examples of a file transfer tool
% adapted to grid infrastructures, providing for instance support for
% parallel streams, authentication, checkpoint/restart in case of
% failures, etc. Based on such tools, catalogue-based data localization
% and management services have been built, such as RLS~\cite{DunGaiGisetal02DataGrid},
% Reptor~\cite{KunLauStoSto05FGCS}, Optor~\cite{KunLauStoSto05FGCS}, LDR~\cite{ldr}. Such
% catalogues allow the user to manually register and characterize data
% copies, but do not provide however any support for transparent access,
% nor for automatic consistency maintenance. Another approach relies on
% the concept of logistical network (such as IBP~\cite{BasBecFagetal02IBP}), which
% gives access to a set of distributed buffers available on the network,
% that can be used for intermediate file storage. NeST~\cite{BenVenLeRetal02}
% proposes a similar approach, enhanced with a mechanism which
% dynamically selects the transport protocol and its parameters. In all
% these systems, the user still has to \emph{explicitly} transfer data
% to/from these buffers and, again, no mechanism is provided for
% maintaining the consistency of replicated data.

However, a few grid data management systems acknowledge that providing a
transparent data access model is important by integrating this idea at
the early stages of their design. 
\emph{Grid file systems}, for instance, provide a
familiar, file-oriented API allowing to transparently access
physically distributed data through globally unique, logical file
paths. The applications simply open and access such files as if they
were stored on a local file system. A very large distributed storage
space is thus made available to those existing applications that usually use
file storage, with no need for modifications. This approach has been
taken by a few projects like GFarm~\cite{TatMorMatetal02Gfarm},
GridNFS~\cite{HonAdaKee05GridNFS}, LegionFS~\cite{WhiWalHumGri01LegionFS}, etc. 

On the other hand, the transparent data
access model is equally defended by the concept of 
\emph{grid data-sharing service}~\cite{AntBerCarDesBouJanMonSen06GDS},
illustrated by the JuxMem platform~\cite{AntBouJan05SCPE}. Such a
service provides the grid applications with the abstraction
of a globally shared memory, in which data can be easily stored and
accessed through global identifiers.  
To meet this goal, the design of
JuxMem leverages the strengths of several building blocks:
consistency protocols inspired by DSM systems; algorithms for
fault-tolerant distributed systems; protocols for scalability and
volatility support from peer-to-peer (P2P) systems. Note that such a
system is fundamentally different from traditional DSM systems 
(such as TreadMarks, etc.).
First, it targets a larger scale through hierarchical
consistency protocols suitable for an efficient exploitation of grids
made of a federation of clusters. 
Second, it addresses from the very beginning the
problem of resource volatility due to failures or to the lack of
resource availability. 

Compared to the grid file system approach, this
approach improves \emph{access efficiency} by totally relying on main
memory storage. Besides the fact that a main memory access is more
efficient than a disk access, the system can leverage
locality-optimization schemes developed for the DSM consistency
protocols.

\paragraph{Limitations.}

However, the JuxMem grid data-sharing service suffers from some
limitations with respect to the efficient storage and access of
massive data blocks. Actually, data are not fragmented in JuxMem: each
individual data is physically stored 
as a single data block in the main memory of
a storage provider, and possibly replicated as such on multiple backup 
providers. Consequently, the largest data block that the service is
able to store is limited by the size of the RAM of a \emph{single}
provider, typically, a few gigabytes. 
This lack of
fragmentation has another drawback regarding load balancing as all
accesses to \emph{different} parts of the same massive block are
served by the \emph{same} RAM provider.

% (e.g. of the order of several tera-bytes). As an example of such
% applications we could consider very large databases whose
% implementation rely on a single file to store the whole data. A
% second limitation of the \JuxMem approach comes from the granularity
% of data accesses: a lock is associated to each data and clients need
% to acquire this lock before any access. Concurrent write accesses to
% different parts of the same data block are thus serialized, which
% may be a serious source of inefficiency when modifying only small
% parts of massive blocks. This does not hold for read operations,
% since \JuxMem allows concurrent read operations to proceed in
% parallel on the same data block. However a bottleneck may be created
% due to a possibly heavy load on the RAM providers that store that
% data block, whatever the grain of the access. Again, this is due to
% the lack of fragmentation: all accesses to \emph{different} parts of
% the same massive block are served by the \emph{same} RAM providers.

Recently, the efficient allocation and access of massive data blocks
in main memory has been addressed by the JumboMem~\cite{PakJoh07JumboMem}
system.  This system is designed for clusters, not for grids. It allows
users to manipulate large contiguous data blocks (of the order of
1~TB) using the aggregated RAM of a set of nodes interconnected
through a high-speed Infiniband System Area
Network. However, JumboMem is targeted for a \emph{single} user and
does not enable data sharing: it does not provide synchronization, nor
replication, nor optimized mechanisms for distributed access by
\emph{multiple} users. In contrast,  a lot of applications in the field of
databases and data-mining target \emph{multi-user} environments. This
requires adding an efficient  concurrency control, which is
not natively provided by JumboMem.

% Applications like database and data mining however target multi-user
% environments, but such an approach needs an efficient underlying
% concurrency control.  For example, most DB systems have adopted
% snapshot isolation (cite) to allow better performance than
% serializability. Multi versioning is particularly adept at
% implementing true snapshot isolation, something which other methods
% of concurrency control frequently do either incompletely or with
% high performance costs.  Thus, support for multi-versioning is
% interesting in our context as well.

\paragraph{Our approach.}

Our contribution is twofold. First, we propose a data sharing service
allowing to store \emph{massive} blocks of data in a distributed,
multi-user environment. Second, \emph{efficient fine-grain access} to
the data is provided thanks to distributed, RAM-based storage of data
fragments, while leveraging a DHT-based metadata
management scheme, which is natively parallel.

This paper is organized as follows. Section~\ref{sec:contrib} gives an
overview of our architecture and describes how data access operations
are handled. Section~\ref{sec:impl} provides a few implementation
details and reports on a preliminary experimental evaluation. 
Finally, on-going
and future work is discussed in Section~\ref{sec:conclusion}.

%%%%%%%%%%%%%%%%%%%%%%%%%%%%%%%%%%%%%%%%%%%%%%%%%%%%%%%%%%%%%%%%%%%%%%

\section{Enabling efficient fine-grain access}
\label{sec:contrib}

Our goal is to provide efficient fine-grain access
to massive data blocks stored in large-scale distributed environments
such as grids.
To goal is addressed in the following way. Data is fragmented into
small equally-sized chunks (which will be called \emph{pages} below)
and distributed across the local memory of a large number of grid
nodes, which act as providers of storage space. This fragmentation
allows: 1) to store huge data blocks; and 2) to avoid contention for
disjoint accesses to pages. To each data block, we associate some
metadata allowing to identify and localize the pages that belong to
that block. In order to avoid contention for metadata access, metadata
is structured in a fine-grained manner to be described below, and stored in
a distributed hash table (DHT).  Finally, efficient large-scale
concurrency both for reads and writes is achieved using \emph{versioning}:
concurrent writes to the same page can proceed in parallel on
multiple versions of that page. Our contribution lies in the
adequate combination of these techniques to achieve efficient fine-grained
access to massive data.

\subsection{Architecture}

Our service relies on a set of distributed processes communicating
through remote procedure calls (RPCs).  In a typical setting, each
process is running on a different physical node.
\begin{description}

\item[Data providers] are responsible for storing and retrieving individual
  pages in their local RAM.

\item[A versioning manager] is responsible for serializing write
  requests and for directing read requests to the latest version 
  available for reading.

\item[Metadata providers] are responsible for storing information
  about the identity and localization of the individual pages that
  make up a data block. In our design, metadata providers are
  organized as a Distributed Hash Table (DHT). Details are given in
  Section~\ref{sec:metadata}.

\item[A provider manager] receives and solves the clients' requests
  for data providers. Available providers must previously register
  with this entity. 

%   In our design, the role of provider manager is
%   also mapped to the DHT mentioned above.

\end{description}
To interact with the service, client processes simply use a client
library, to which they pass a list of DHT gateways and the network id (IP
address, port) of the versioning manager. The rest of the system is
transparent to the clients.

\subsection{User interface}
\label{sec:API}

Clients manipulate massive data blocks through a simple API:
\begin{lstlisting}
  block_id = alloc(page_size, data_size)
  block_version = write(block_id, local_buffer, offset, size)
  read(block_id, block_version, local_buffer, offset, size)
\end{lstlisting}
Massive blocks are identified and accessed through a globally unique
id, generated when the block is allocated. The user is able to control
the granularity (\lstinline{page_size}) and maximal size of the block
(\lstinline{data_size}). Fine-grain access for reads and writes is enabled
through $(\Offset, \Size)$ range queries. \emph{Each write generates a new
block version.} Read operations may explicitly reference a block
version. By default, they return the latest available version.

% Reads and writes obey sequential consistency semantics. New
% snapshots versions (\emph{new\_snap}) are generated incrementally.

\subsection{Metadata organization}
\label{sec:metadata}

Metadata serves the purpose of identifying and localizing the pages
corresponding to the range $(\Offset, \Size)$ specified by read and
write operations. Our design aims at favoring fast concurrent accesses
to metadata.

\begin{figure}
  \centerline{%
    \includegraphics[width=.4\textwidth]{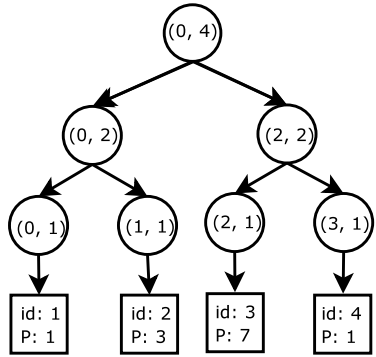}%
  }%
  \caption{Metadata representation for a 4-page block. Leaves store
    page ids \myblock{Id} and the corresponding provider ids
    \myblock{P}.
    All nodes are
    labeled with the $(\Offset,\Size)$ range they cover.}%
  \label{fig:metadata}
\end{figure}

When the user allocates $\Datasize$ bytes for a block, the service
actually allocates $\Adjustedsize$ bytes, where $\Adjustedsize$ is
the smallest power of~2 larger than $\Datasize$. We organize metadata
as a full binary tree. At each level, the nodes of the tree cover disjoint $(\Offset,
\Size)$ ranges. The root covers $(0, \Adjustedsize)$, that is, the whole
data block.  An intermediate node covering $(\Offset, \Size)$ points to its
left child covering $(\Offset, \Size / 2)$, and to its right child
covering $(\Offset + \Size / 2, \Size / 2)$. Leaves cover single pages
and point to the page id and to provider holding the page (see
Figure~\ref{fig:metadata}).

A tree node covering $(\Offset, \Size)$ is identified by a \emph{key},
obtained by applying a hashing function on the tuple $(\Blockid,
\Offset, \Size, \Blockversion)$.  Intermediate tree nodes store the
following information: $\Offset$, $\Size$, $\Leftkey$ and $\Rightkey$,
which are respectively the keys of its left and right child. Leaves
(covering single pages) store a $\Pageid$ and a $\Providerid$.  Tree
nodes are stored on the metadata providers using a DHT structure using
the keys defined above. This approach is inspired by Merkle
trees~\cite{Mer88Signature}, initially developed to handle Lamport's one-time
signatures.

By relying on the DHT architecture and by selecting an adequate hashing
function, an even distribution of page requests among metadata
providers can be guaranteed with a high probability. Each client takes
profit of this even distribution by simultaneously contacting a large
number of different gateways to the DHT service when executing parallel
requests.

\subsection{Managing allocs, reads and writes}
\label{sec:how}

\begin{figure}
  \centerline{
    \hfill
    \subfigure[Reading a block: sequence of RPC calls]%
    {\includegraphics[width=.45\textwidth]{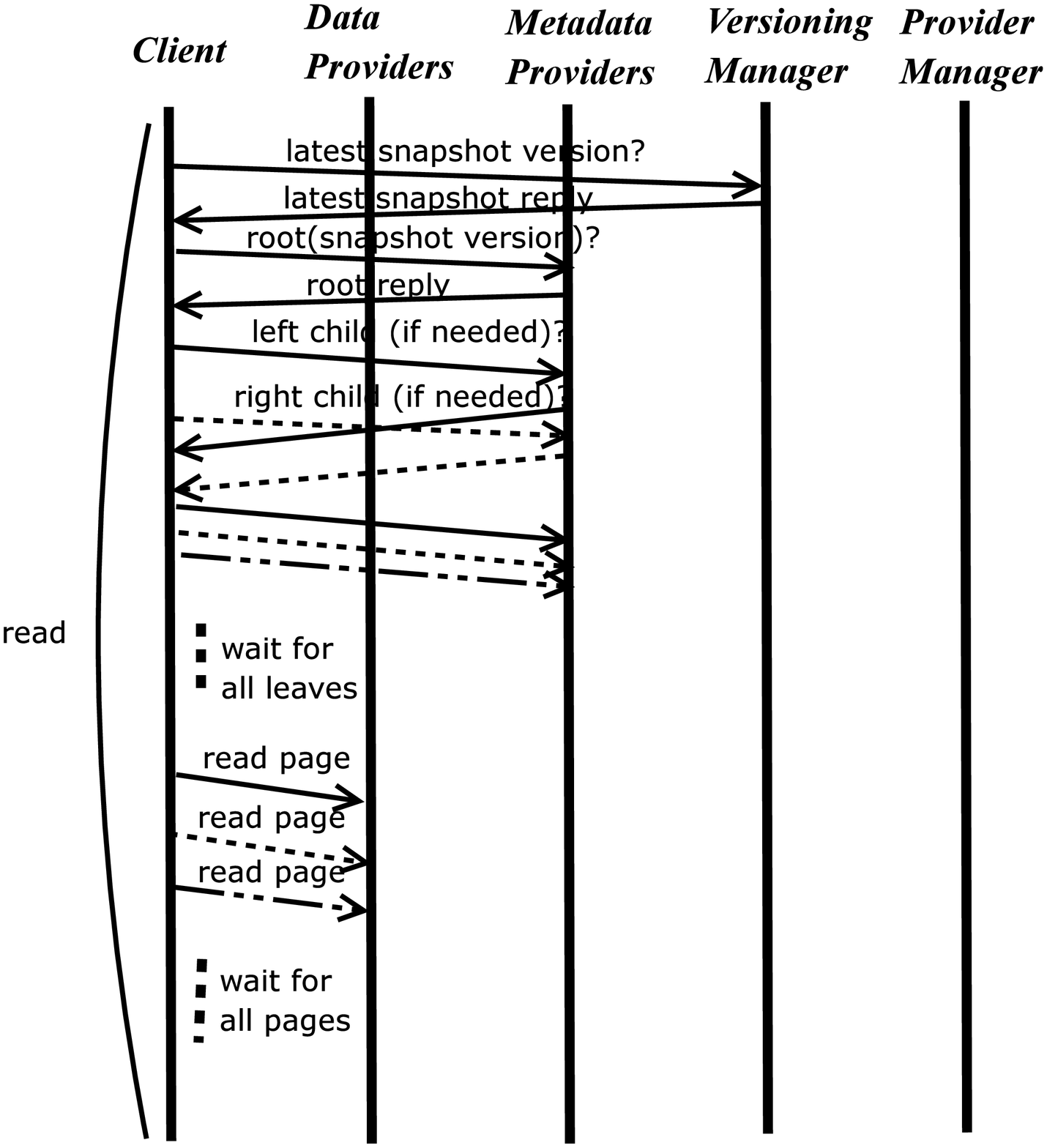}
      \label{fig:rpc_reads}}%
    \hfill
    \subfigure[Writes: sequence of RPC calls]%
    {\includegraphics[width=.45\textwidth]{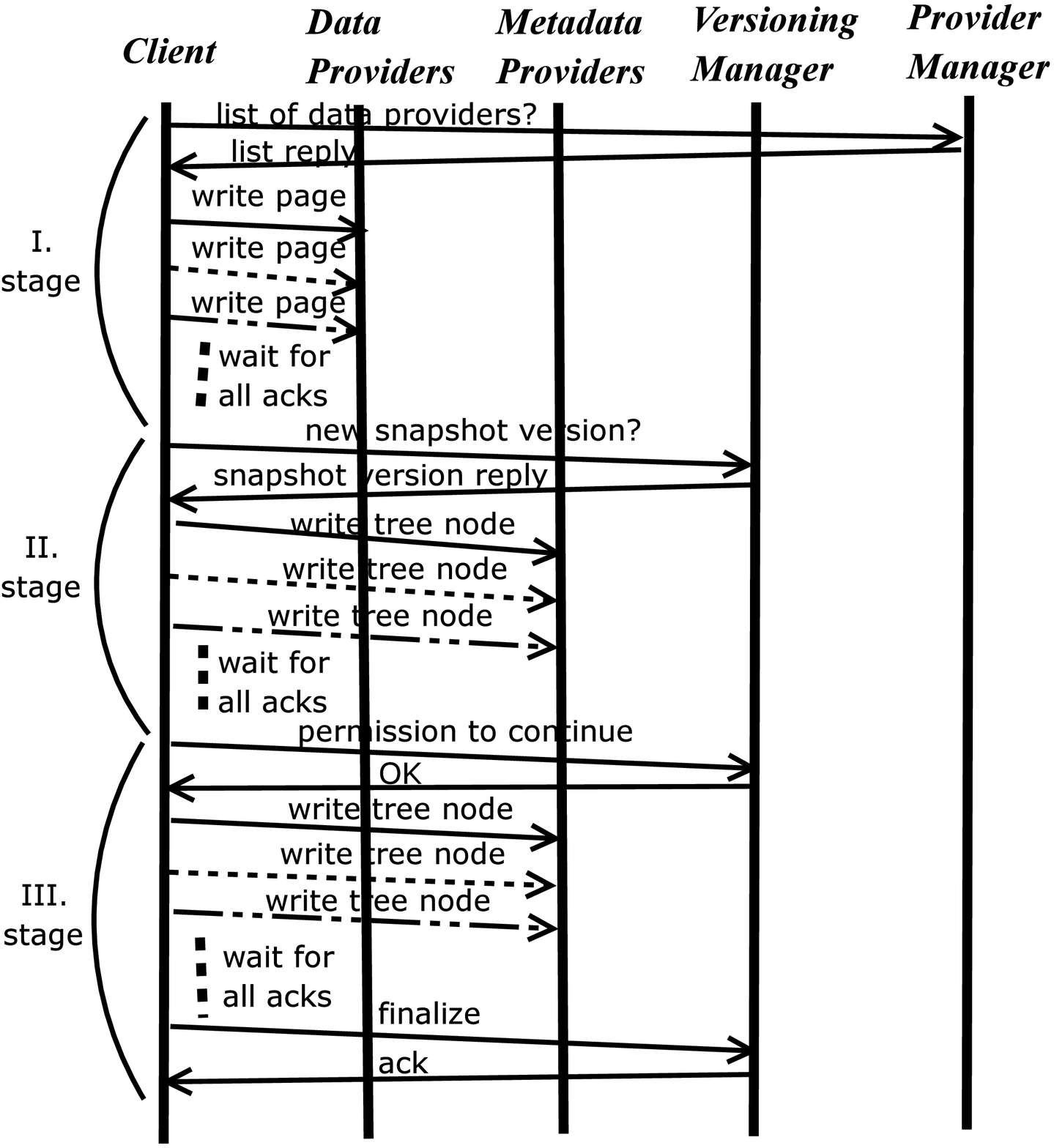}
      \label{fig:rpc_writes}}%
    \hfill
  }
  \caption{Managing reads and writes: different line styles denote
    RPCs running in parallel}
  \label{fig:Reads and writes}
\end{figure}

Allocation is the cheapest and simplest operation. The client merely
contacts the versioning manager providing a page size and total block
size. The versioning manager assigns this block an initial version number, $0$.

% and marks the block ready to accept requests, returning success.  No
% further nodes are contacted.

To perform a read if no block version is specified, the client
(Figure~\ref{fig:rpc_reads}) contacts the versioning manager and requests
the latest block version available. If a block version is specified by the
read operation, then this step is simply skipped. Then, the client contacts
the metadata providers and recursively queries  the tree nodes
covering the range given by $(\Offset, \Size)$ for that particular block
version, starting from the root and descending towards the leaves. When
a leaf is reached, the client directly contacts the appointed data provider
and downloads the actual page. The read operation completes
successfully when all the pages have been downloaded. It fails if a node or
a page could not be retrieved.  In order to enhance
parallelism, requests and responses for tree nodes and for pages are
handled asynchronously by multiple threads on the client side, and are
served in parallel by the various metadata and data providers,
respectively.

% How about ranges larger than the buffer size? ;-)

A write operation (Figure~\ref{fig:rpc_writes}) initiated by the client completes in
several stages.
\begin{enumerate}
\item
The client contacts the provider manager to retrieve a list of active data
providers available to store the pages in $(\Offset, \Size)$ to be
written. 
After receiving the reply, it associates a random data
provider and a random page id to each page, so as to uniquely 
identify the page in the
system with high probability. Then, it contacts the data providers,
requesting them to store the pages. 
As in the case of the read operation, write
requests sent to providers are asynchronously handled by the client,
and served in parallel by the data providers.

\item
After all providers acknowledge that the pages have been stored, 
the client contacts the versioning manager to
receive a new version number which shall identify the new block version. The
versioning manager enqueues this write request, marks it as pending
and returns the version number to the client. After receiving it,
the client generates the corresponding tree nodes with respect to the
new block version, starting from the leaves up to this new root.  All
tree nodes whose range is totally included in the interval $[\Offset,
\Offset + \Size]$ are written to the metadata providers. The rest of the
nodes are stored for later processing. The goal of this processing is
to properly handle concurrent metadata updates for a single block.

\item 
Then, the client contacts the versioning
manager requesting permission to complete the write operation. If this
write request is the oldest one in the queue, then the versioning manager
grants permission to complete the write. Otherwise it waits
for previous pending writes to be dequeued before granting
permission. After receiving permission, the client builds the
remaining tree nodes. These nodes cover ranges not included in
$[\Offset, \Offset + \Size]$. They must correctly reference their children
corresponding to nodes not modified by the current write, by using the
latest block version previously completed. At the end of this stage,
all generated tree nodes are sent to the metadata providers.

\item
Finally, the client confirms write completion to the versioning
manager, which dequeues the write and marks its corresponding version
as the latest block version.

\end{enumerate}

Note that various versions of the same page may be stored on
different providers: for each new page version to be written, the
least loaded known provider is chosen for its storage, in order to
preserve a global load balance in terms of amount of data stored by
the providers. (The precise description of this scheme is out of the scope of
this paper.)

An important consequence of this property is that
successive incremental versions of a data block can be stored as
long as storage space is still \emph{globally} available in the
system: thanks to our choice of preserving a global load balance, a
provider will run out of storage space only when \emph{all} the providers
collectively reach their storage limits. In this case, ad-hoc garbage
collection can be used to remove the oldest version of the data
block. Such a feature has not been implemented in our system, yet.

\section{Implementation and experimental evaluation}
\label{sec:impl}

Evaluations are performed using the Grid'5000~\cite{Grid5000} testbed,
a reconfigurable, controllable and monitorable experimental Grid
platform spread over 9~sites geographically distributed in France.  We
use 160~nodes of a Grid'5000 cluster. Each node has a Intel Pentium~4
CPU running at 2.6~GHz under Linux~2.6 (Ubuntu), outfitted with 4~GB
of RAM each, and interconnected by a Gigabit Ethernet network. The
theoretical maximum network bandwidth is thus 125~MB/s. However, if we
consider the IP and TCP header overhead, this maximum becomes slightly
lower: 117.5~MB/s for a MTU of 1500~B. In practice, we could measure
111~MB/s for a standard TCP socket end-to-end transfer.

\subsection{Implementation details}

We use BambooDHT~\cite{RheGodKaretal05Bamboo}, which provides a stable, scalable DHT
implementation on top of which we build the abstraction of our
metadata providers and of the provider manager.

The providers and the versioning manager are implemented in C++ using
the Boost C++ collection of libraries. We chose Boost for its
standardization throughout the C++ community, and for the wide range
of functionalities it provides, among which serialization, threading
and asynchronous I/O are of particular interest to us.

\subsection{Performance and evaluation}

In this section, we assess the effectiveness of our implementation by
running a set of experiments. To support our claim of efficiently
dealing with both massive blocks and fine-grain access, we fix the
allocated block size at 1~TB, and the page size at 64~kB for all
our experiments. Thus, the metadata tree  generates a significant
overhead
% (corresponding to 1~TB)
as the actual data accesses will
concern various continuous ranges from 16~MB up to 1~GB within the
overall range of 1~TB.

\paragraph{Using a single client.}

\begin{figure}[p]
  \centering
  \subfigure[Cost of read accesses when using 5, 50, 100 metadata
  providers]%
  {\includegraphics[width=0.85\textwidth]{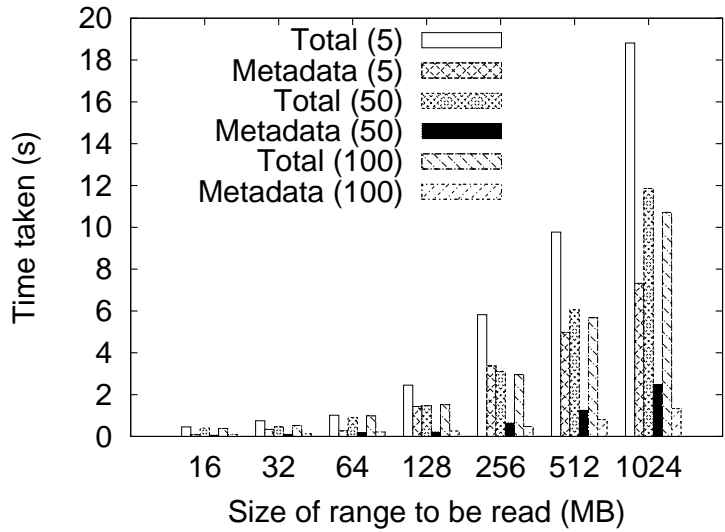}
    \label{fig:single_read}}
  \\ \leavevmode \vfill \leavevmode \\
  \subfigure[Cost of write accesses when using 5, 50, 100 metadata
  providers]%
  {\includegraphics[width=0.85\textwidth]{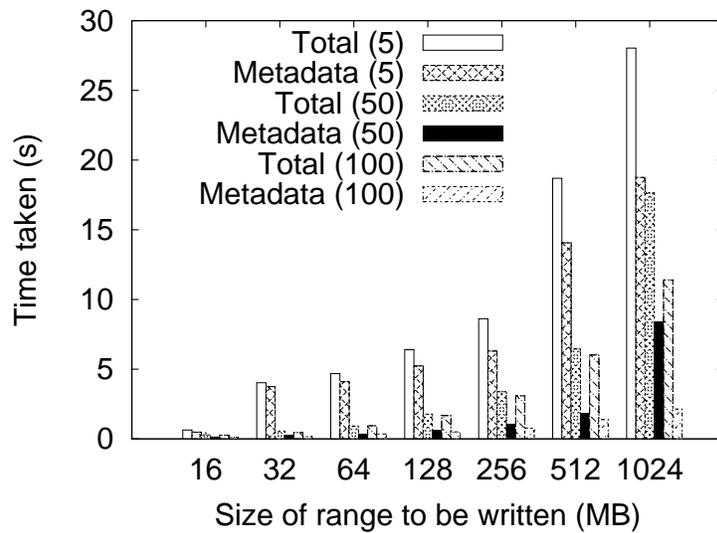}%
    \label{fig:single_write}}
  \caption{Average data access cost for various contiguous segment sizes
    and a variable number of metadata providers. For each number of
    metadata providers, we sort out the time needed to manage the metadata.}
  \label{fig:single_results}
\end{figure}

Our first series of experiments (Figure~\ref{fig:single_results})
assesses the overhead of metadata management.  We deploy one
versioning manager, 100~data providers, and a variable number of
metadata providers.
Each physical node runs at most one data provider and one metadata provider.
A single client writes a series of data, and then reads them back.

It first writes a range \Size of 16~MB starting from \Offset~0. 
Then, it continues writing a second range of 32~MB, starting
from the end of the previous range, and so on, 
doubling the \Size parameter each time until
writing a range of~1~GB. Then, the client successively reads back each
of the consecutive segments. 

The individual writing and reading times
for each segment are logged, sorting out the time used in managing the
metadata with respect to the total writing or reading time. Such a
cycle is repeated 100~times. This experiment is done for several
numbers of metadata providers, that is, several sizes of the DHT,
ranging from~5 to~100. 

The average timings are reported on
Figure~\ref{fig:single_results}. 
Of course, the larger the DHT, the larger the
degree of parallelism in accessing its nodes from the client's threads,
whence the shorter the overall time.

These timings show an overhead of 18\,\% for metadata read operations
(Figure~\ref{fig:single_read}) and 23\,\% for metadata write operations
(Figure~\ref{fig:single_write}) for 100~metadata providers. This effectively
results in a bandwidth of 92~MB/s for reads and 86~MB/s for writes
in accessing the final 1~GB range, 
to be compared to the maximal limit of 111~MB/s measured in
standard TCP socket end-to-end transfer.  On the other hand, 
using only 5~metadata providers results in a
metadata management overhead exceeding 68\%,
which demonstrates the benefits of using a large number of metadata
providers, that is, a large DHT.

\paragraph{Using multiple concurrent clients.}

\begin{figure}
  \centerline{
    \hfill
    \subfigure[Bandwidth for reads]%
    {\includegraphics[width=0.45\textwidth]{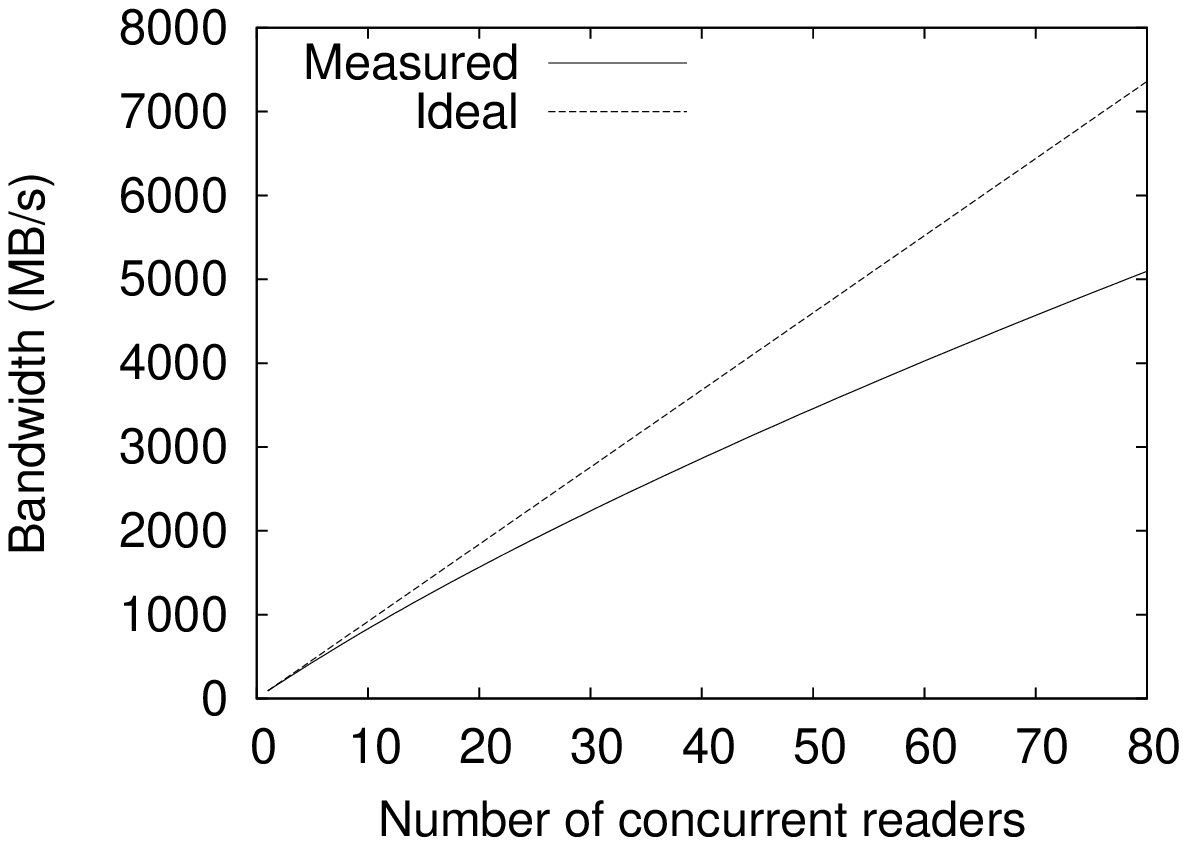}
      \label{fig:multi_read}}
    \hfill
    \subfigure[Bandwidth for writes]%
    {\includegraphics[width=0.45\textwidth]{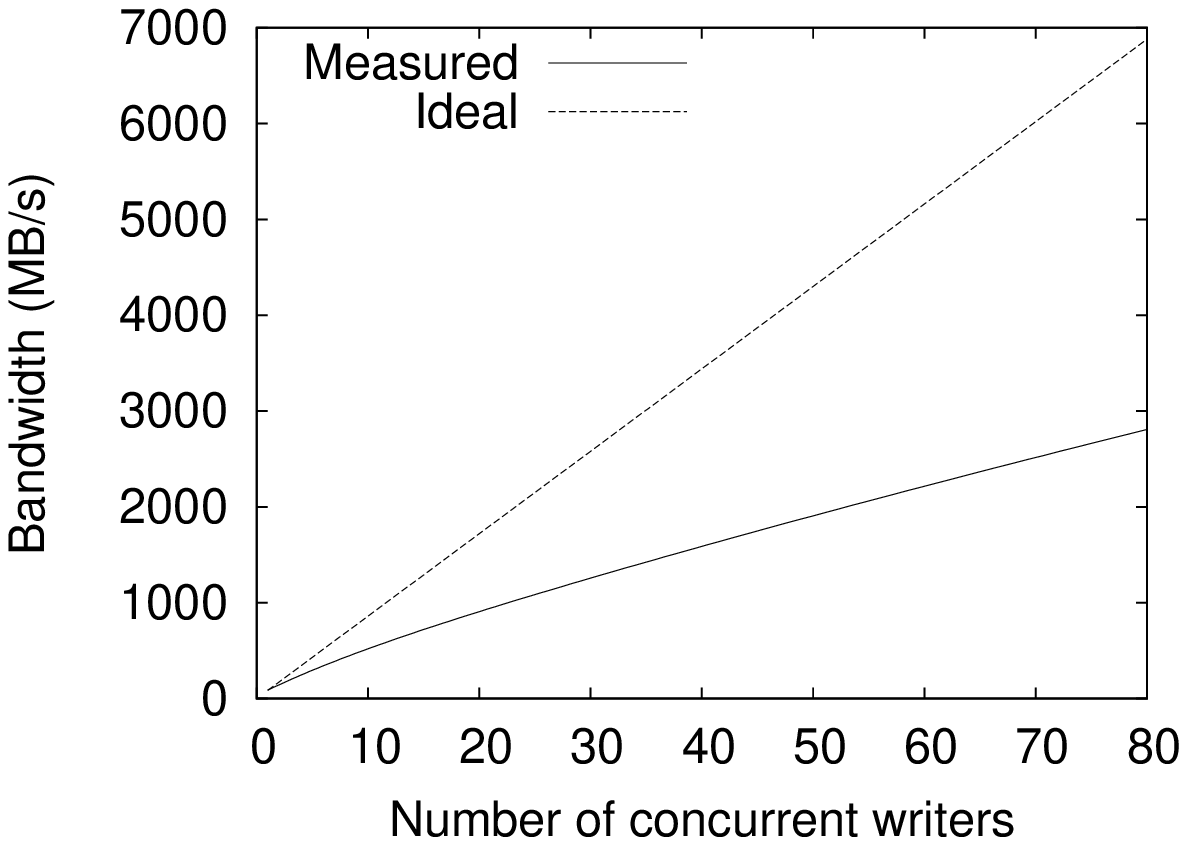}
      \label{fig:multi_write}}
    \hfill
  }
  \caption{Average aggregated bandwidth when varying the number of concurrent
    clients}
  \label{fig:multi_results}
\end{figure}

Our next series of experiments (Figure~\ref{fig:multi_results})
benchmarks our system in a highly concurrent environment, evaluating
its scalability when increasing the number of simultaneous reads and
writes. For comparison, we also report on what we call an ``ideal''
bandwidth corresponding to the aggregation of totally independent read
(resp., write) operations. That is, we multiply the bandwidth
of a single reader (resp., writer) by the number of readers
(resp., writers).  

We deploy 80~data providers and 80~metadata
providers.
Each physical node runs one data provider and one metadata
provider. The versioning manager is run on a separate node.
Then, we deploy a variable number of
clients, each of which being run on a separate node, different
from the ones used for data, metadata and versioning manager. Clients
are synchronized to start simultaneously.  They either read or write a
disjoint range of the block: client $i$ uses 
$\Offset = i \times 64~\MB$, $\Size =
64~\MB$.  For reads, data is prewritten. We measure the average
aggregated bandwidth, both for reads and writes, and compare it to the
ideal aggregation of bandwidth obtained from a single reader/writer.

As it can be observed, the fine-grained dispersion of data and metadata
allows for high bandwidth under heavy concurrency, especially for
reads (Figure~\ref{fig:multi_read}). Writes suffer from a slight
performance penalty because of metadata synchronization
(Section~\ref{sec:how}). Contacting different providers and different metadata
providers concurrently enables a high degree of load balancing among
the network nodes. As such, it makes up for the metadata overhead
observed in the first series of experiments.

\section{Conclusion}
\label{sec:conclusion}

We have addressed the problem of efficiently storing massive data of
the order of terabytes in a grid distributed environment. Our contribution
consists in proposing a data-sharing service allowing to efficiently
allocate, access and modify such massive blocks of data in a distributed,
multi-user environment. Efficient fine-grain access to arbitrarily
small parts of the data is provided thanks to distributed, RAM-based
storage of data fragments, while leveraging a DHT-based, natively 
parallel metadata management scheme. Preliminary experiments performed
with our prototype using the Grid'5000 testbed show that our approach
scales well, both in terms of storage providers and in terms of
concurrency degree.

% Considering resources, in our case main memory for grid nodes, are
% very cheap and in large quantities it is convenient to consider
% sacrificing space for speed.

Our prototype is however a work in progress and definitely demands further
refinement. Fault tolerance, which becomes critical in grid
environments, is only partially addressed. We currently leverage some
fault-tolerance mechanisms provided by the DHT on which we rely for
the implementation of some of the entities of our architecture, the
metadata providers and the provider manager. This enhances the
availability of metadata thanks to the underlying replication used by
the DHT. However, the versioning manager, though not under heavy load,
is still a single point of failure in this preliminary
scheme. Besides, data is not replicated: for each page, a single copy
is kept on a single provider. In order to improve fault-tolerance,
replication-based mechanisms could be envisioned in both cases. To this
purpose, we intend to explore the possibility to use self-organizing
groups to represent these entities, built on fault-tolerant
distributed algorithms for atomic multicast, as
in~\cite{AntDevMon06CPE}.

While targeting database, data-mining and multimedia applications, we
have not experimented, yet, with a standard implementation that could
use our service. We are considering interfacing our service with the
PostgreSQL DBMS, in order to provide an efficient support for snapshot
isolation.

% providing efficient transaction management by optimizing snapshot
% isolation for use with our system.

%%\begin{mysmall}
\bibliographystyle{splncs_with_empty_author_in_misc}
\bibliography{paper,juxmem}
%%\end{mysmall}

\end{document}